\documentclass[
    ,final            
  ]
  {aipproc}

\layoutstyle{6x9}

\def\be{\begin{equation}}
\def\ee{\end{equation}}
\def\bea{\begin{eqnarray}}
\def\eea{\end{eqnarray}}
\def\beq{\begin{eqnarray}}
\def\eeq{\end{eqnarray}}

\def\nn{\nonumber \\}
\def\e{{\rm e}}

\begin{document}

\title{Modified gravity as realistic candidate for dark energy, inflation
and dark matter\footnote{
Based on the talk given at DSU08 conference, June 2008.
}}



\author{Shin'ichi Nojiri}{
  address={Department of Physics, Nagoya University, Nagoya 464-8602, Japan},
}
\author{Sergei D. Odintsov}{
  address={Instituci\`{o} Catalana de Recerca i Estudis Avan\c{c}ats (ICREA)
and Institut de Ciencies de l'Espai (IEEC-CSIC),
Campus UAB, Facultat de Ciencies, Torre C5-Par-2a pl, E-08193 Bellaterra
(Barcelona), Spain},
}

\begin{abstract}
We review the unification of early-time inflation with late-time
acceleration in several local modified gravity models which pass Solar
System and cosmological tests. It is also demonstrated that account of
non-local gravitational corrections to the action does not destroy the
possibility of such unification. Dark matter effect is caused by composite
graviton degree of freedom in such models.

\end{abstract}

\maketitle

Modified gravity (for general review of different models, see
\cite{review}) is known to be
very interesting candidate for dark energy. From another side, there
exist modified gravity theories which may naturally unify the early-time
inflation and late-time acceleration \cite{prd}. Recently, there were
suggested realistic theories \cite{prd1,prd2} which pass the local as well
as cosmological
tests and unify inflation with late-time acceleration (with the presence
of radiation/matter dominance epoch at intermediate universe). We will
review such models which represent $F(R)$ gravity or its combination with
non-local gravity below.

We will start from the theory with the action:
\be
\label{Uf0}
S= \int d^4 x \sqrt{-g} \left[ \frac{1}{2\kappa^2}\left( R + f(R) \right) + L_m \right]\ .
\ee
Here $f(R)$ is an appropriate function of the scalar curvature $R$ and $L_m$ 
is the Lagrangian density of matter.
In order to generate the inflation, one may require
\be
\label{Uf1}
\lim_{R\to\infty} f (R) = - \Lambda_i\ .
\ee
Here $\Lambda_i$ is an effective cosmological constant at the early universe and therefore
it is natural to assume $\Lambda_i \gg \left(10^{-33}{\rm eV}\right)^2$.
For instance, it could be $\Lambda_i\sim 10^{20\sim38}\left( {\rm eV}\right)^2$.
In order that the current cosmic acceleration could be generated,
let us consider that currently $f(R)$ is a small constant, that is,
\be
\label{Uf3}
f(R_0)= - 2\tilde R_0\ ,\quad f'(R_0)\sim 0\ .
\ee
Here $R_0$ is the current curvature $R_0\sim \left(10^{-33}{\rm eV}\right)^2$.
Note that $R_0> \tilde R_0$ due to the contribution from matter. In fact, if
we can regard $f(R_0)$ as an effective cosmological constant, the effective
Einstein equation gives $R_0=\tilde R_0 - \kappa^2 T_{\rm matter}$.
Here $T_{\rm matter}$ is the trace of the matter energy-momentum tensor.
We should note that $f'(R_0)$ need not vanish exactly. Since we are considering
the time scale of one-ten billion years, we only require
$\left| f'(R_0) \right| \ll \left(10^{-33}\,{\rm eV}\right)^4$.
The last condition is
\be
\label{Uf4}
\lim_{R\to 0} f(R) = 0\ ,
\ee
which means that there is a flat spacetime solution.

Instead of the model corresponding to (\ref{Uf1}), we may consider a model which satisfies
\be
\label{UU2}
\lim_{R\to\infty} f (R) = \alpha R^m \ ,
\ee
with a positive integer $m>1$ and a constant $\alpha$.
In order to avoid the anti-gravity $f'(R)>-1$, we find $\alpha>0$ and
therefore $f(R)$ should be positive at the early universe.
On the other hand, Eq.(\ref{Uf3}) shows that $f(R)$ is negative at
the present universe. Therefore $f(R)$ should cross zero in the past.

At the early universe,  $f(R)$-term in (\ref{UU2})  dominates the
Einstein-Hilbert
term. Now let assume that there exists matter with an equation of state parameter $w$.
For spatially flat FRW universe, the scale factor $a(t)$ behaves as
\be
\label{UUU2}
a(t) \propto t^{h_0}\ ,\quad h_0 \equiv \frac{2m}{3(w+1)}\ .
\ee
Then the effective equation of the state parameter $w_{\rm eff}$, which is defined by
\be
\label{UUU3}
w_{\rm eff}= -1 + \frac{2}{3h_0}\ ,
\ee
can be less than $-1/3$ and the accelerating expansion could occur if $m$ is large enough although
$w_{\rm eff} > -1$, which is quintessence type. Then the inflation could
occur due to the $R^m$ behavior of $f(R)$ in (\ref{UU2}).

We may consider a model \cite{prd1}
\be
\label{UU2d}
f(R)= \frac{\alpha R^{2n} - \beta R^n}{1 + \gamma R^n}\ .
\ee
Here $\alpha$, $\beta$, and $\gamma$ are positive constants and $n$ is a positive integer.
(\ref{UU2d}) gives \cite{prd1}
\be
\label{UUU7}
R_0=\left\{ \left(\frac{1}{\gamma}\right)
\left(1+ \sqrt{ 1 + \frac{\beta\gamma}{\alpha} }\right)\right\}^{1/n}\ ,
\ee
and therefore
\be
\label{UU6}
f(R_0) \sim -2 \tilde R_0 = \frac{\alpha}{\gamma^2}
\left( 1 + \frac{\left(1 - \frac{\beta\gamma}{\alpha} \right)
\sqrt{ 1 + \frac{\beta\gamma}{\alpha}}}{2 + \sqrt{ 1 + \frac{\beta\gamma}{\alpha}}} \right) \ .
\ee
Then we find
\be
\label{UU9}
\alpha \sim 2 \tilde R_0 R_0^{-2n}\ ,\quad \beta \sim 4 {\tilde R_0}^2 R_0^{-2n} R_I^{n-1}\ ,\quad
\gamma \sim 2 \tilde R_0 R_0^{-2n} R_I^{n-1}\ .
\ee
In general $F(R)$ gravity contains graviton and scalar as physical modes. The propagation of
the scalar field might give a large correction to the Newton law.
In the model (\ref{UU2d}), the correction to the Newton law could be small since the mass $m_\sigma$
of the extra scalar filed is large and given by
$m_\sigma^2 \sim 10^{-160 + 109 n}\,{\rm eV}^2$ and in the air on the earth,
$m_\sigma^2 \sim 10^{-144 + 98 n}\,{\rm eV}^2$ \cite{prd1}.
In both cases, the mass $m_\sigma$ is very large if $n\geq 2$.

As a model corresponding to (\ref{Uf1}), we may consider\cite{prd2}
\bea
\label{tan7}
f(R) &=& -\alpha_0 \left( \tanh \left(\frac{b_0\left(R-R_0\right)}{2}\right)
+ \tanh \left(\frac{b_0 R_0}{2}\right)\right) \nn
&& -\alpha_I \left( \tanh \left(\frac{b_I\left(R-R_I\right)}{2}\right)
+ \tanh \left(\frac{b_I R_I}{2}\right)\right)\ .
\eea
We now assume
\be
\label{tan8}
R_I\gg R_0\ ,\quad \alpha_I \gg \alpha_0\ ,\quad b_I \ll b_0\ ,
\ee
and
\be
\label{tan8b}
b_I R_I \gg 1\ .
\ee
When $R\to 0$ or $R\ll R_0,\, R_I$, $f(R)$ behaves as
\be
\label{tan9}
f(R) \to - \left(\frac{\alpha_0 b_0 }{2\cosh^2 \left(\frac{b_0 R_0}{2}\right) }
+ \frac{\alpha_I b_I }{2\cosh^2 \left(\frac{b_I R_I}{2}\right) }\right)R\ .
\ee
and $f(0)=0$ again. When $R\gg R_I$, it follows
\bea
\label{tan10}
f(R) \to - 2\Lambda_I &\equiv&
 -\alpha_0 \left( 1 + \tanh \left(\frac{b_0 R_0}{2}\right)\right)
 -\alpha_I \left( 1 + \tanh \left(\frac{b_I R_I}{2}\right)\right) \nn
&\sim& -\alpha_I \left( 1 + \tanh \left(\frac{b_I R_I}{2}\right)\right)\ .
\eea
On the other hand, when $R_0\ll R \ll R_I$, one gets
\bea
\label{tan11}
f(R) &\to& -\alpha_0 \left[ 1 + \tanh \left(\frac{b_0 R_0}{2}\right)\right]
 - \frac{\alpha_I b_I R}{2\cosh^2 \left(\frac{b_I R_I}{2}\right) } \nn
&\sim& -2\Lambda_0 \equiv -\alpha_0 \left[ 1 + \tanh \left(\frac{b_0 R_0}{2}\right)\right] \ .
\eea
Here we have assumed (\ref{tan8b}). We also find
\be
\label{tan12}
f'(R)= - \frac{\alpha_0 b_0 }{2\cosh^2 \left(\frac{b_0 \left(R - R_0\right)}{2}\right) }
- \frac{\alpha_I b_I }{2\cosh^2 \left(\frac{b_I \left(R - R_I\right)}{2}\right) }\ ,
\ee
which has two valleys when $R\sim R_0$ or $R\sim R_I$. When $R= R_0$, we obtain
\be
\label{tan13}
f'(R_0)= - \alpha_0 b_0 - \frac{\alpha_I b_I }{2\cosh^2 \left(\frac{b_I \left(R_0 - R_I\right)}{2}\right) }
> - \alpha_I b_I - \alpha_0 b_0 \ .
\ee
On the other hand, when $R=R_I$, we get
\be
\label{tan14}
f'(R_I)= - \alpha_I b_I - \frac{\alpha_0 b_0 }{2\cosh^2 
\left(\frac{b_0 \left(R_0 - R_I\right)}{2}\right) }
> - \alpha_I b_I - \alpha_0 b_0 \ .
\ee
Then, in order to avoid the anti-gravity period, one obtains
\be
\label{tan15}
\alpha_I b_I + \alpha_0 b_0 < 2\ .
\ee
We now investigate the correction to the Newton law and the matter instability issue.
In the solar system domain, on or inside the earth, where $R\gg R_0$,
$f(R)$ can be approximated by 
\be
\label{tan16}
f(R) \sim -2 \Lambda_{\rm eff} + 2\alpha \e^{-b(R-R_0)}\, .
\ee
On the other hand, since $R_0\ll R \ll R_I$, by assuming Eq.~(\ref{tan8b}), $f(R)$ in
(\ref{tan7}) could be also approximated by
\be
\label{tan17}
f(R) \sim -2 \Lambda_0 + 2\alpha \e^{-b_0(R-R_0)}\, ,
\ee
which has the same expression, after having identified $\Lambda_0 = \Lambda_{\rm eff}$ and $b_0=b$.
Then, we may check the case of (\ref{tan16}) only.
We find that the effective mass has the following form
\be
\label{tan18}
m_\sigma^2 \sim \frac{\e^{b(R-R_0)}}{4\alpha b^2}\, ,
\ee
which could be very large, that is, $m_\sigma^2 \sim 10^{1,000}\,{\rm eV}^2$
in the solar system and $m_\sigma^2 \sim 10^{10,000,000,000}\,{\rm eV}^2$ in the air
surrounding the earth, and the correction to Newton's law can be made negligible.
Hence, both models pass local tests and may unify the inflation with dark
energy. Moreover, the arguments presented in ref.\cite{dm} show that such
models may also describe dark matter consistently. The reason is caused by
composite graviton degree of freedom. Thus, not only modification of
gravitational potential occurs (this gravitational correction explains the
galaxy rotation curves) but also such gravitational dark matter shows the
particles-like properties as predicted by observational data.

As third model we consider the non-local
gravity \cite{non1,non2,koivisto,non3,non4}, whose action
is given by
\be
\label{nl1}
S=\int d^4 x \sqrt{-g}\left\{
\frac{1}{2\kappa^2}R\left(1 + f(\left(\nabla^2\right)^{-1}R)\right) + {\cal L}_{\rm matter}
\right\}\ .
\ee
Here $f$ is some function and $\nabla^2$ is the d'Almbertian for scalar field.
The above action can be rewritten by introducing two scalar fields $\phi$
and $\xi$ in the following form \cite{non2}:
\be
\label{nl2}
S=\int d^4 x \sqrt{-g}\left[
\frac{1}{2\kappa^2}\left\{R\left(1 + f(\phi)\right)
 - \partial_\mu \xi \partial^\mu \phi - \xi R \right\}
+ {\cal L}_{\rm matter}
\right]
\ .
\ee
By the variation over $\xi$, we obtain
$\nabla^2\phi=R$ or $\phi=\left(\nabla^2\right)^{-1}R$.
Substituting the above equation into (\ref{nl2}), one re-obtains (\ref{nl1}).

Especially in case
\be
\label{NLdS2}
f(\phi)=f_0 \e^{b\phi}= f_0 \e^{-2H_0 \phi}\ ,
\ee
we obtain a de Sitter solution:
\be
\label{NLdS1}
H=H_0\ ,\quad
\phi= - 4H_0 t\ ,\quad
\xi= - \frac{3f_0}{3 - 4b} \e^{-4bH_0 t} + \frac{\xi_0}{3H_0}\e^{-3H_0 t} +1\ .
\ee
We may discuss the accelerating early-time and late-time cosmology in the
non-local gravity where $F(R)$-term is added (the appearance of local
and non-local corrections is typical for string
theory low-energy effective action). The starting action is: 
\be
\label{nl30}
S=\int d^4 x \sqrt{-g}\left\{
\frac{1}{2\kappa^2}R\left(1 + f(\left(\nabla^2\right)^{-1}R)\right) 
+ F(R) + {\cal L}_{\rm matter}
\right\}\ .
\ee
Here $F(R)$ is some function of $R$.

We may propose several scenarios. One is that the inflation at the early
universe is generated mainly by $F(R)$ part but the current acceleration is defined
mainly by $f\left(\left(\nabla^2\right)^{-1}R\right)$ part. One may consider the inverse,
that is, the inflation is generated by
$f\left(\left(\nabla^2\right)^{-1}R\right)$ part but the late-time acceleration by $F(R)$.

For instance, for the first scenario one can take: $F(R)= \beta R^2$.
Here $\beta$ is a constant. We choose $f(\left(\nabla^2\right)^{-1}R)$ part as in
(\ref{NLdS2}) but $f_0$ is taken to be very small
and $\phi$ starts with $\phi=0$. Hence, at the early universe
$f\left(\left(\nabla^2\right)^{-1}R\right)$ is very small and
could be neglected. Then due to the $F(R)$-term,
there occurs (slightly modified) $R^2$-inflation. After the end of the inflation,
there occurs the radiation/matter dominance era. In this phase, $f(\phi)$ becomes
large as time goes by and finally this term dominates. As a result, deSitter expansion
occurs at the present universe.

For the second scenario, the early-time inflation
is generated by $f\left(\left(\nabla^2\right)^{-1}R\right)$ part 
but the cosmic acceleration is generated
by $F(R)$. As an $F(R)$-term, one can take the model \cite{HS}:
\be
\label{HS1}
F_{HS}(R)=-\frac{m^2 c_1 \left(R/m^2\right)^n}{c_2 \left(R/m^2\right)^n + 1}\ ,
\ee

The estimation of ref.\cite{HS} suggests that $R/m^2$ is not so small but rather large even
at the present universe and $R/m^2\sim 41$.
Hence, $F_{HS}(R)\sim - \frac{m^2 c_1}{c_2} + \frac{m^2 c_1}{c_2^2}
\left(\frac{R}{m^2}\right)^{-n}$, which gives an ``effective'' cosmological
constant $-m^2 c_1/c_2$ and generates the late-time accelerating expansion.
One can show that
\be
\label{HSbb1}
H^2 \sim \frac{m^2 c_1 \kappa^2 }{c_2} \sim \left(70 \rm{km/s\cdot pc}\right)^2
\sim \left(10^{-33}{\rm eV}\right)^2\ .
\ee
At the intermediate epoch, where the matter density $\rho$ is larger
than the effective cosmological constant, $\rho > \frac{m^2 c_1}{c_2}$,
there appears the matter dominated phase and the universe
expands with deceleration. Hence, above model describes the effective
$\Lambda$CDM cosmology.

As a $f\left(\left(\nabla^2\right)^{-1}R\right)$ part, we consider theory (\ref{NLdS2}) with $b=1/2$,
again. It is assumed $f_0$ is large and $f\left(\left(\nabla^2\right)^{-1}R\right)$ term could be
dominant at the early universe.
Hence, following (\ref{NLdS1}), $\phi$ becomes negative and large as time
goes by and therefore
$f(\phi)$ becomes small and could be neglected at late universe. Then
there appears naturally the radiation/matter dominated phase.
After that due to $F_{HS}(R)$-term (\ref{HS1}),
the late-time acceleration occurs.

Thus, the possibility of consistent description of the universe expansion
history from the early-time inflation till late-time acceleration in
modified gravity (without/with non-local term) is demonstrated.
As related gravitational phenomenon the appearance of dark matter is
naturally explained in such models.


\begin{theacknowledgments}
The work by S.D.O. was
supported in part by MEC (Spain) projects FIS2006-02842 and
PIE2007-50I023, RFBR grant 06-01-00609 and LRSS project N.2553.2008.2.
The work by S.N. is supported in part by the Ministry of Education,
Science, Sports and Culture of Japan under grant no.18549001 and Global
COE Program of Nagoya University provided by the Japan Society
for the Promotion of Science (G07).

\end{theacknowledgments}


\begin{thebibliography}{9}

\bibitem{review}
S.~Nojiri and S.~D.~Odintsov,
arXiv:hep-th/0601213.

\bibitem{prd}
S.~Nojiri and S.~D.~Odintsov,
Phys.\ Rev.\ D {\bf 68}, 123512 (2003)
[arXiv:hep-th 0307288].

\bibitem{prd1}
S.~Nojiri and S.~D.~Odintsov,
Phys.\ Rev.\ D {\bf 77}, 026007 (2008)
[arXiv:0710.1738[hep-th]].

\bibitem{prd2}
G.~Cognola, E.~Elizalde, S.~Nojiri, S.~D.~Odintsov, L.~Sebastiani
and S.~Zerbini,
Phys.\ Rev.\ D {\bf 77}, 046009 (2008)
[arXiv:0712.4017[hep-th]].

\bibitem{dm}
S.~Nojiri and S.~D.~Odintsov,
arXiv:0807.0685[hep-th];
arXiv:0801.4843[astro-ph].

\bibitem{HS}
W.~Hu and I.~Sawicki,
arXiv:0705.1158;
Y.~Song, H.~Peiris and W.~Hu,
arXiv:0706.2399;
S.~Capozziello, M.~De~Laurentis, S.~Nojiri and S.~D.~Odintsov,
arXiv:0808.1335[hep-th].


\bibitem{non1}
S.~Deser and R.~Woodard,
Phys.\ Rev.\ Lett. {\bf 99}, 111301 (2007).


\bibitem{non2}
  S.~Nojiri and S.~D.~Odintsov,
  Phys.\ Lett.\  B {\bf 659}, 821 (2008)
  [arXiv:0708.0924 [hep-th]].

\bibitem{koivisto}
T.~Koivisto, Phys.\ Rev.\ D {\bf 77}, 123513 (2008);
[arXiv:0807.3778[gr-qc]].


\bibitem{non3}
  S.~Jhingan, S.~Nojiri, S.~D.~Odintsov, M.~Sami, I.~Thongkool and S.~Zerbini,
  Phys.\ Lett.\  B {\bf 663}, 424 (2008)
  [arXiv:0803.2613 [hep-th]].
\bibitem{non4}
S. Capozziello, E. Elizalde, S. Nojiri and
S.D. Odintsov, [arXiv:0809.1535[hep-th]].


\end{thebibliography}
\end{document}